\documentstyle[preprint,aps]{revtex}
\begin{document}
\title{Theory of AC Josephson Effect in Superconducting Constrictions }
\author{A.V. Zaitsev}
\address{Institute of Radioengineering and Electronics, Russian\\
Academy of Sciences, 107903, Moscow, Russia}
\author{D.V. Averin}
\address{Department of Physics, State University of New York, \\
Stony Brook, NY 11794}
\maketitle

\begin{abstract}
We have developed a microscopic theory of ac Josephson effect in short
ballistic superconducting constrictions with arbitrary electron
transparency, and constrictions with diffusive electron transport. As
applications of the theory we study smearing of the subgap current
singularities by pair-breaking effects in the superconducting electrodes of
the constriction, and the structure of these singularities in constrictions
between the composite S/N electrodes with the proximity-induced gap in the
normal layer.
\end{abstract}


\narrowtext

It is quite surprising that despite the long history of Josephson junctions,
the microscopic theory of the ac Josephson effect in junctions with
arbitrary transparency is still absent. It is known that the basic mechanism
of electron transport in such junctions at finite bias voltages is the
process of multiple Andreev reflections (MAR) \cite{BTK}, but the first
attempts at quantitative description of MAR based on the Green's function
method \cite{AVZ,Z,Z1,Ar} did not produce final results. Final progress in
this direction has been made recently \cite{GZ,AB1} but only for the case of
fully transparent junctions. Quantitative description of MAR in junctions
with arbitrary transparency is possible within the framework of the
Bogolyubov-de Gennes (BdG) equations \cite{BSW,AB,Sp,AB2}, but this approach
does not allow to incorporate inelastic scattering and pair-breaking effects
in the junction electrodes. This circumstance is crucial because dynamics of
MAR makes finite-voltage electron transport in high-transparency junctions
very sensitive to such scattering processes. The aim of the present work was
to develop the theory of MAR in junctions with arbitrary transparency
between the superconductors with general microscopic structure. The theory
allows us to study, for example, the smearing of the MAR-related current
singularities by pair-breaking scattering, and the ac Josephson effect
between the normal conductors with the proximity-induced superconducting
order parameter.

The basic model of a high-transparency Josephson junction is a short
superconducting constriction (shorter than the coherence length $\xi$ and
elastic and inelastic scattering lengths of the electrodes) with a
transparency $D$. The general approach to the description of such a
constriction requires that we use the Green's function technique. In this
technique, the constriction is described with the non-equilibrium
quasiclassical Green's function $\check{G}$ which is 4x4 matrix consisting
of 2x2 retarded, advanced, and Keldysh matrixes $\hat{G}^{R,A}$ and $\hat{G}$
\cite{LO,Z1}. To calculate the current we need to know only the asymmetric
part $\check{J}$ of the Green's function $\check{J}=\check{G}(p_{zj}) -%
\check{G}(-p_{zj})$. Here $p_z$ is momentum in the transport direction, and $%
j=1,2$ numbers the constriction electrodes. Solving the quasiclassical
equations for $\check{G}$ inside the two electrodes and matching the
solution across the constriction with the boundary conditions \cite{Z1} one
can show that the matrix $\check{J}(t, t^{\prime })$ is continuous inside
the constriction and is given by the expression \cite{Z1}:
\begin{equation}
\check{J}=2D\check{G}_{-}*\check{G}_{+}*({\bf \check{1}-}D \check{G}_{-}*%
\check{G}_{-})^{-1}=\check{G}_{-}*(\check{G}_{+}+ i\lambda {\bf \check{1}}%
)^{-1}-(\check{G}_{+}+i\lambda {\bf \check{1}})^{-1}*\check{G}_{-} \, .
\label{J}
\end{equation}
Here the product denoted by $*$ means the convolution with respect to the
internal time\ variable, i.e., $\check{G}_\mu *\check{G}_\eta =\int dt_1
\check{G}_\mu (t,t_1)\check{G}_\eta (t_1,t^{\prime })$, and we used the
following notations: $\lambda =\sqrt{(1-D)/D},\;{\bf \check{1}=}\check{1}%
(t-t^{\prime }),\;\check{G}_{\pm }=(\check{G}_1\pm \check{G}_2)/2,\;\check{G}%
_j(t,t^{\prime })=\check{S}_j(t){\it \check{g}}_j(t-t^{\prime })\check{S}%
_j^{*}(t^{\prime })$. In the last equation, ${\it \check{g}}_j(t)=\int
\check{g}_j(\epsilon )\exp (-i\epsilon t)d\epsilon /(2\pi )$ is the
equilibrium Green's function of $j$th superconductor: $\hat{g}_j (\epsilon)=[%
\hat{g}_j^R(\epsilon )-\hat{g}_j^A(\epsilon )]\tanh (\epsilon /2T),\;\hat{g}%
_j^R(\epsilon )=g_j^R(\epsilon )\hat{\tau}_z+ f_j^R(\epsilon )i\hat{\tau}%
_y=-[\hat{g}_j^A(\epsilon )]^{*}$, the matrix $\check{S}_j$ is : $\check{S}%
_j(t)=\exp [i\varphi _j(t)\check{\tau}_z/2]\;$where $\varphi _j(t)$ is the
phase of the order parameter in the $j$th electrode, and the phase
difference $\varphi=\varphi _2-\varphi _1$\ is determined by the applied
voltage $V(t)$: $\stackrel{.}{\varphi }(t)=(2e/\hbar )V(t)$. Equation (\ref
{J}) gives the following expressions for the Keldysh component of $\check{J}$%
:
\begin{eqnarray}
\hat{J} &=&(\hat{G}_2^R*\hat{Q}_1\;-\hat{Q}_1*\hat{G}_2^A-\hat{G}_1^R* \hat{Q%
}_2+\hat{Q}_2*\hat{G}_1^A)/2  \label{JK} \\
&=&\hat{J}^R*{\it \hat{f}}_{+}-{\it \hat{f}}_{+}*\hat{J}^A+ (\hat{G}_2^R*%
\hat{P}_1-\hat{P}_1*\hat{G}_2^A-\hat{G}_1^R*\hat{P}_2+ \hat{P}_2*\hat{G}%
_1^A)/2  \label{JK1}
\end{eqnarray}
where $\hat{Q}_j=\hat{Q}^R*\hat{G}_j*\hat{Q}^A,\;\hat{Q}^\mu =(\hat{G}
_{+}^\mu +i\lambda {\bf \hat{1}})^{-1},\;\hat{J}^\mu =\hat{G}_{-}^\mu *\hat{Q%
}^\mu -\hat{Q}^\mu *\hat{G}_{-}^\mu ,\;\hat{P}_j=\hat{Q}^R* (\hat{G}_j^\mu *%
{\it \hat{f}}_{-}-{\it \hat{f}}_{-}*\hat{G}_j^A)* \hat{Q}^A,\;{\it \hat{f}}%
_{\pm }=({\it \hat{f}}_1\pm {\it \hat{f}}_2 )/2$, and ${\it \hat{f}}%
_j(t,t^{\prime})=\hat{S}_j(t){\it f}(t- t^{\prime }) \hat{S}_j^{*}(t^{\prime
})$ is the matrix distribution function of the $j$th superconductor with $%
{\it f}(\tau)=\int \tanh (\epsilon /2T)\exp (-i\epsilon \tau )d\epsilon
/(2\pi )$. The current in the constriction can be found from eq.\ (\ref{JK})
or (\ref{JK1}):
\begin{equation}
I(t)=\frac \pi {2eR_0}\mbox{Tr}\hat{\tau}_z\hat{J}(t,t)\, ,  \label{Iball}
\end{equation}
where $R_0=\pi \hbar /e^2$ for a single-mode constriction. For a
constriction with large number of propagating electron modes and
cross-section area ${\cal A}$, angular averaging over momentum direction
should be taken into account in (\ref{Iball}), and $R_0=(2\pi
/ep_{F1})^2\hbar ^3/{\cal A}$, where $p_{F1}= \min p_{Fj}$.

Equations (\ref{JK}), (\ref{JK1}) can be used to find the current not only
in the ballistic constrictions but also in short diffusive constrictions,
i.e. channels with large number of propagating modes and length $d$ that
satisfies the condition $l\ll d\ll \xi $, where $l$ is elastic scattering
length. Indeed, in this case solution of the quasiclassical equations for
Green's function gives for the matrix $\check{J}$ the expression $\check{J}%
=(2lv_{Fz}/dv_F)\ln (\check{G}_2*\check{G}_1)$ \cite{AVZ}, which reduces to
the following form\ \cite{Z2}:
\begin{equation}
\check{J}=\frac{4lv_{Fz}}{dv_F}\int_0^\infty \frac{d\lambda }{\sqrt{
1+\lambda ^2}}\left[ \check{G}_{-}*(\check{G}_{+}+i\lambda {\bf \check{1}}%
)^{-1}-(\check{G}_{+}+i\lambda {\bf \check{1}})^{-1}* \check{G}_{-}\right] .
\label{Jd}
\end{equation}
After the substitution: $\lambda =\sqrt{(1-D)/D}$, we get for the current
\begin{equation}
I(t)=\frac \pi {4eR_N}\int_0^1\frac{dD}{D\sqrt{1-D}}\mbox{Tr} \hat{\tau}_z
\hat{J}(t,t;D)\, ,  \label{Idiff}
\end{equation}
where $\hat{J}(t,t;D)$ is given by eq.\ (\ref{JK}), and $R_N$ is the
normal-state resistance of the channel. Equation (\ref{Idiff}) shows that
similarly to the approach based on the BdG equations \cite{AB2}, in the
general Green's function approach, the current in the diffusive
superconducting constriction can be written as a sum of independent
contributions from infinite number of ballistic propagating modes with the
distribution of transparencies given by the Dorokhov's \cite{Dor} density
function $(\pi\hbar /2e^2R_N)/D\sqrt{1-D}$.

To calculate the current at arbitrary bias voltages across the constriction
we should transform eqs.\ (\ref{JK}) and (\ref{JK1}) further. Separating the
normal-state contribution $V/R_N$, where $R_N=R_0/D$ is the normal-state
junction resistance, we obtain the following symmetrized expression for the
current:
\begin{equation}
I(t)=\frac{V(t)}{R_N}+\delta I_{12}(t)+\delta I_{21}(t)\, ,  \label{C}
\end{equation}
\begin{equation}
\delta I_{jk}(t)=\frac \pi {4eR_0}\mbox{Tr}[\hat{\tau}_z(\hat{G}_j^R *\hat{q}%
_{jk}-\hat{q}_{jk}*\hat{G}_j^A)](t,t) \, ,  \label{C1}
\end{equation}
where $\hat{q}_{jk}=\hat{q}_{jk}^R*{\it \hat{g}}_k*\hat{q}_{jk}^A,\; \hat{q}
_{jk}^\mu =2/(\hat{G}_j^\mu +{\it \hat{g}}_k^\mu +2i\lambda {\bf \hat{1}}%
)^{-1}$, and ${\it \hat{g}}_k={\it \hat{g}}_k^R*{\it f}- {\it f}*{\it \hat{g}%
}_k^A$. In equation (\ref{C1}), all distribution functions correspond to
zero potential, and in contrast to the previous definition, the function $%
\hat{G}_1^\mu$ and $\hat{G}_2^\mu$ are defined similarly, $\hat{G}_j^\mu =%
\hat{S} (t){\it \hat{g}}_j^\mu (t-t^{\prime })\hat{S}^{*}(t^{\prime })$,
where $\hat{S}(t)=\exp [i\varphi (t)\hat{\tau}_z/2]$. Different
representations for the current can be obtained from eq.\ (\ref{C1}). We
will use the representation in which the Fourier components $\hat{q}%
_{jk}^{R,A}(\epsilon ,\epsilon ^{\prime })$\ are expressed, respectively,
through the matrices $\hat{\alpha}_{jk}^R(\epsilon , \epsilon ^{\prime })$
and $\hat{\alpha}_{jk}^A(\epsilon ,\epsilon ^{\prime })=[\hat{\alpha}%
_{jk}^R(\epsilon ^{\prime },\epsilon )] ^{\dagger }$, where, for example, $%
\hat{\alpha}_{jk}^R$ obeys the following equation:
\begin{equation}
(\hat{\rho}_k^R{\bf -}\hat{\Gamma}_j^R*\hat{\eta}_k^R)* \hat{\alpha}_{jk}^R=r%
{\bf \hat{1}+}\hat{\Gamma}_j^R \, ,  \label{Syst}
\end{equation}
In this equation, $r=\sqrt{1-D}$ is the reflection amplitude of the
constriction, and other notations are: $\rho _k^R(t,t^{\prime })= {\bf \hat{1%
}}-r\Gamma _k^R(t-t^{\prime })\hat{\tau}_x,\;\hat{\eta}_k^R (t,t^{\prime
})=-r{\bf \hat{1}}+\Gamma _k^R(t-t^{\prime }) \hat{\tau}_x,\;\Gamma
_k^R(t)=\int \gamma _k^R(\epsilon )\exp (-i \epsilon t)d\epsilon /(2\pi ),\;%
\hat{\Gamma}_k^R(t,t^{\prime })= \Gamma _k^R(t-t^{\prime })\exp \{i\hat{\tau}%
_z[\varphi (t)+\varphi (t^{\prime })]/2\}\hat{\tau}_x,\;\gamma _k^R(\epsilon
)=f_k^R(\epsilon )/[g_k^R(\epsilon )+1],\;\gamma _k^A(\epsilon )=\gamma
_k^{R*}(\epsilon )$. As we will see later, $\gamma _k^R(\epsilon )$ has the
meaning of amplitude of Andreev reflection at the interface between the
constriction and the $k$th superconductor. In terms of $\hat{\alpha}%
_{jk}^{R(A)}$ eq. (\ref{C1}) reduces to the following form:
\begin{equation}
\delta I_{jk}(t)=\frac 1{8eR_0}\int \frac{d\omega }{2\pi }\int d\epsilon
J_{jk}(\epsilon ,\epsilon -\omega )e^{-i\omega t} \, ,  \label{dJ}
\end{equation}
where
\[
J_{jk}(\epsilon ,\epsilon ^{\prime })=\mbox{Tr}i\hat{\tau}_y[\gamma
_k^R(\epsilon )W_k(\epsilon ^{\prime })\hat{\alpha}_{jk}^R(\epsilon
,\epsilon ^{\prime })-\gamma _k^A(\epsilon ^{\prime })W_k(\epsilon ) \hat{%
\alpha}_{jk}^A(\epsilon ,\epsilon ^{\prime })]
\]
\[
+ \lbrack 1+\gamma _k^R(\epsilon )\gamma _k^A(\epsilon ^{\prime })] \mbox{Tr}
\hat{\tau}_z \int \frac{d\epsilon _1}{2\pi }W_k(\epsilon _1)\hat{\alpha}%
_{jk}^R (\epsilon ,\epsilon _1)\hat{\alpha}_{jk}^A(\epsilon _1,\epsilon
^{\prime }),\;\;W_k(\epsilon )\equiv \left[ 1-\left| \gamma _k^R (\epsilon )
\right|^2\right] {\it f}(\epsilon ).
\]
Equations (\ref{dJ}) and (\ref{C}) show that the problem of finding the
current in short ballistic superconducting constriction for arbitrary
time-dependent bias voltage reduces to the problem of solving eq. (\ref
{Syst}), which is a Fredgholm integral equation (see, e.g., \cite{MF}).
Similar procedure applied to (\ref{Idiff}) gives the current in diffusive
constriction. It should be noted that this approach assumes that all
frequencies (frequency of Josephson oscillations and typical frequency of
voltage variations) are much smaller than the inverse of the time of
electron motion through the constriction.

Equation (\ref{Syst}) can be solved easily in the case of the dc bias
voltage $V$, when the phase difference is: $\varphi (t)=\omega _Jt+
\varphi_0 $, where $\omega _J=2eV/\hbar $ is the Josephson oscillation
frequency. In this case solution of eq. (\ref{Syst}) can be written as a
series\ , $\hat{\alpha}_{jk}(\epsilon ,\epsilon ^{\prime })=2\pi \sum_n\hat{%
\alpha}_{n(jk)}(\epsilon ^{\prime })\delta (\epsilon - \epsilon ^{\prime
}-n\hbar\omega _J)$ (we omit the superscript R), in which the amplitudes $%
\hat{\alpha}_{n(jk)}$ are determined by a system of recurrence relations. To
obtain these relations explicitly it is convenient to write the matrix $\hat{%
\alpha}_{jk}$ as
\[
\hat{\alpha}_{jk}=\left(
\begin{array}{cc}
b^{+} & a^{-} \\
a^{+} & b^{-}
\end{array}
\right) _{(jk)} \, .
\]
Equation (\ref{Syst}) shows that the pairs of functions $a_{jk}^s,\; b_{jk}^s
$ with $s=\pm $ satisfy the same equations but corresponding to different
polarity of the bias voltage. For dc voltage we get the following recurrence
relations for the matrix elements of $\hat {\alpha}_{n(jk)}$:
\begin{eqnarray}
a_{n+1}{\bf -}\gamma _j(\epsilon _{2n+1})\gamma _k(\epsilon
_{2n})a_n+r[\gamma _j(\epsilon _{2n+1})b_n-\gamma _k(\epsilon
_{2n+2})b_{n+1}] &=&\gamma _j(\epsilon _1)\delta _{n0},  \label{RS} \\
c(\epsilon _{2n+1})b_{n+1}-d(\epsilon _{2n})b_n+c(\epsilon _{2n-1}) b_{n-1}
&=&-r\delta _{n0} \, ,  \nonumber
\end{eqnarray}
where we used the simplified notations $a_n\equiv a_{n(jk)}^{+} (\epsilon )$
and $b_n\equiv b_{n(jk)}^{+}(\epsilon )$, and other notations are: $\epsilon
_n=\epsilon +neV,\;c(\epsilon )=D\gamma _j (\epsilon )\gamma_k(\epsilon
+eV)/[1-\gamma _j^2(\epsilon )],\; d(\epsilon )=1-\gamma_k^2(\epsilon
)+D\gamma _j^2(\epsilon +eV)/ [1-\gamma _j^2(\epsilon+eV)]+D\gamma
_k^2(\epsilon )[1-\gamma _j^2 (\epsilon -eV)]$. The amplitudes $a_{n(jk)}\;$%
and\ $b_{n(jk)}$ determine Fourier components of the current $%
I(t)=\sum_nI_n\exp (in\omega _Jt)$, which according to eqs.\ (\ref{C}) and (%
\ref{dJ}) are given by the expression $I_n=V\delta_{n,0}/R_N+I_{n(12)}+
I_{n(21)}$, where
\begin{equation}
I_{n(jk)}=\frac 1{2eR_0}\int_{-\infty }^\infty d\epsilon W_k (\epsilon
)\Phi_{n(jk)}(\epsilon ),  \label{I3}
\end{equation}
\[
\Phi _{n(jk)}(\epsilon )=-\gamma _k(\epsilon _{2n})a_{n(jk)}
(\epsilon)-\gamma _k^{*}(\epsilon _{-2n})a_{-n(jk)}^{*}(\epsilon )+
\]
\[
\sum_m[1+\gamma _k(\epsilon _{2m+2n})\gamma _k^{*}(\epsilon
_{2m})][b_{n+m}b_m^{*}-a_{n+m}a_m^{*}]_{(jk)}(\epsilon ), .
\]
For a constriction between two BCS superconductors the recurrence relations (%
\ref{RS}) and equation (\ref{I3}) for the current reduce to the
corresponding expressions that can be obtained from the BdG equations - see
\cite{AB}, where these expressions were derived for symmetric constriction.
This means that for the purpose of description of the ac Josephson effect in
a short constriction all information about the microscopic structure of a
superconducting electrode of the constriction is contained in the function $%
\gamma ^R(\epsilon)$ (introduced after eq. (\ref{Syst})) which has the
meaning of the amplitude of Andreev reflection from this superconductor.

Equations (\ref{RS}) can be solved by standard methods \cite{MF} (see also
\cite{BSW}). Namely, it follows from (\ref{RS}) that $b_n
(\epsilon)=b_0(\epsilon )\prod_{m=\pm 1}^np_{\pm }(\epsilon _{2m})$, where
the functions $p_{\pm }(\epsilon )$ are solutions of the equations $%
p_{\pm}(\epsilon )=c(\epsilon _{\mp 1})/[d(\epsilon )- c(\epsilon _{\pm
1})p_{\pm}(\epsilon _{\pm 2})]$; the two signs ($\pm $) here correspond to\ $%
n\geq 1\;(n\leq -1)$, and $b_0(\epsilon )=r[d(\epsilon )-c(\epsilon
_1)p_{+}(\epsilon )-c(\epsilon _{-1})p_{-}(\epsilon )]^{-1}$. These
relations provide the basis for convenient numerical evaluation of the
current. They also can be used to find analytical solutions in certain
cases. In particular, in the low voltage limit, $eV\ll \Delta _j$, we get
explicitly $p_{\pm }(\epsilon )=\kappa (\epsilon )-\sqrt{\kappa
^2(\epsilon)-1}$\ where$\;\kappa (\epsilon )=\{[1{\bf -}\gamma _1^2
(\epsilon)][1-\gamma _2^2(\epsilon )]+D\gamma _1^2(\epsilon )+D\gamma
_2^2(\epsilon)\}/2D\gamma _1(\epsilon )\gamma _2(\epsilon )$, and $%
b_0(\epsilon)=r/2D\gamma _1(\epsilon )\gamma _2(\epsilon )\sqrt {\kappa
^2(\epsilon )-1}$. Thus, eqs. (\ref{RS}) and (\ref{I3}) enable us to find
the current in the constriction between the superconductors with arbitrary
quasiparticle spectrum. As was discussed above, this feature of our approach
is very important, since all deviations of the quasiparticle spectrum from
its ``ideal'' BCS form affect strongly the subharmonic gap structure and
especially the low-voltage behavior of the current.

The quasiparticle spectrum of a superconductor can differ significantly from
the BCS form, in particular, due to pair-breaking effects. One of the most
important example of these effects is scattering on paramagnetic impurities
-- see, e.g., \cite{Maki,Abr}. As a first application of our general theory
we consider the ac Josephson effect in a constriction between two
superconductors with paramagnetic impurities. The retarded Green's functions
of such superconductors are determined by the relations \cite{Maki}:
\begin{equation}
g^R=\frac u{\sqrt{u^2-1}}=uf^R\,,\;\;\;\;\frac \epsilon \Delta =u [1-\frac
\zeta {\sqrt{1-u^2}}]\, ,  \label{gfr}
\end{equation}
where $\zeta $ is the pair-breaking parameter, $\zeta =\hbar /\tau _s \Delta
$, with $\tau _s$ being the spin-flip scattering time. Similar expressions
for the Green's functions are valid also in the situations with other
pair-breaking effects \cite{Maki}, for instance, for dirty thin
superconducting film of thickness $d_S\ll \xi $ in magnetic field $H$
parallel to the film. The Green's functions of such a film are given by eq.\
(\ref{gfr}) with the pair-breaking parameter $\zeta = lv_F(eHd_S)^2/18\Delta$%
. Weak pair-breaking effects result in smearing of the BCS singularity in
the quasiparticle spectrum and suppression of the superconducting energy gap
to a reduced value $\Delta _g=\Delta (1-\zeta ^{2/3})^{3/2}$. The gap
disappears completely at $\zeta \geq 1$. Figure 1 shows how these changes in
the quasiparticle spectrum are reflected in the current-voltage ($I\!-\!V$)
characteristic of the constriction. All gap-related features in the $I\!-\!V$
characteristic are rapidly broadened at small $\zeta $, and it becomes
practically linear in the gapless regime $\zeta \geq 1$.

As another application of our theory we consider a constriction between two
normal conductors in which superconductivity is induced by the proximity
effect, i.e. an $S/NcN/S$ junction. Besides general interest to the
proximity effect, the importance of this model is due to its relevance for
realistic description of the high critical current tunnel junctions \cite
{AGK}. We consider a particular case of a thin dirty $N$ layer of thickness $%
d_n\ll\xi _n$ with low transparency of the $S/N$ interface $\left\langle
D^{\prime}\right
\rangle \ll 1$. It is assumed that the resistance of the
interface is still negligible in comparison to the constriction resistance.%
\cite{rem} The Green's functions in the $N$ layer are given then by the
first equation in (\ref{gfr}) with $u=(\epsilon +i\gamma _bg_S^R)/i\gamma
_bf_S^R$ -- see, e.g., \cite{AGK,VZ} and references therein. Here $g_S^R$
and $f_S^R$ are the Green's functions of the superconductor, and $\gamma
_b/\hbar =\left
\langle D^{\prime }\right\rangle v_{Fn}/4d_n$ is the
characteristic tunneling rate across the $S/N$ interface which is assumed to
be larger than electron-phonon inelastic scattering rate in the $N$ layer.
The energy gap $\Delta _g$ is induced in the $N$ layer due to the proximity
effect. If the $S$ electrode of the structure is the BCS superconductor with
energy gap $\Delta$, the induced gap $\Delta _g$ is determined by the
equation $\Delta_g=\Delta \gamma _b/[\sqrt{\Delta ^2-\Delta _g^2}+\gamma _b]$%
. Existence of the induced gap implies that there are two peaks in the
density of states of the $N$ layer, at energies $\Delta _g$ and $\Delta $.
This structure of the density of states results in a complex structure of
the subharmonic gap singularities in the $I-V$ characteristics of the
constriction. Example of such a structure is shown in Fig.2 for $\gamma
_b/\Delta =1$, when $\Delta_g\simeq 0.54\Delta$. We see that the most
pronounced current singularities in this case are the subharmonic
singularities at $eV=2\Delta _g/n$ associated with the induced gap. Also
visible are the singularities at ``combination'' energies $\Delta +\Delta _g$
and $(\Delta +\Delta _g)/2$.

In the case of fully transparent constriction with $D=1$ we can obtain
explicit analytical expressions for the current since the recurrence
relations (\ref{RS}) can be solved directly in this case. In particular, in
the low voltage limit $eV\ll \Delta _g{\it f}( \Delta_g)$ we get for
symmetric constrictions:
\begin{equation}
I(t)=\frac 1{2eR_N}\int d\epsilon \mbox{Re}\frac{i\sin \varphi
F_{+}(\epsilon ,V)+2u\sqrt{u^2-1}F_{-}(\epsilon ,V)}{u^2(\epsilon )-
\cos^2(\varphi /2)} \, ,  \label{Isym}
\end{equation}
\[
F_{\pm }(\epsilon ,V)=[F(\epsilon ,V)\pm F(\epsilon ,-V)]/2 \, ,
\]
\[
F(\epsilon ,\pm V)={\it f}(\epsilon )-\int_{\pm \Delta _g}^\epsilon
d\epsilon ^{\prime }\frac{\partial {\it f}(\epsilon ^{\prime })}{ \partial
\epsilon ^{\prime }}\exp \left( -\int_\epsilon ^{\epsilon ^{\prime }}\frac{dE%
}{eV}\ln \left| \gamma ^2(E)\right| \right) \, .
\]
These expressions imply that the current can be written as a sum of two
contributions, one from the two discrete states inside the energy gap with
energies $\epsilon _{\pm }=\pm \epsilon _\varphi $ determined by the
equation $u(\epsilon _\varphi )=\left| \cos (\varphi /2) \right|$, and
another contribution from the continuum of states above the gap. In contrast
to constrictions between the BCS superconductors \cite{FT,BH,AB1}, in the
general case of the non-BCS spectrum of quasiparticles both the discrete and
continuum contributions to the current are significant. In particular, for
the ballistic constriction in presence of pair-breaking effects one finds
from eqs. (\ref{gfr}) and (\ref{Isym}) that the current $I_d$ carried by
the subgap states is: $I_d=\overline{I}\left| \sin (\varphi /2)\right|
\left( 1-\zeta /\left| \sin ^3(\varphi /2)\right| \right) \Theta \left(
\left| \sin (\varphi /2)\right| -\zeta ^{1/3}\right)$, where $\overline{I}=2%
\mbox{sgn}(V)\Delta _g{\it f}(\Delta _g)/eR_N$. At $\zeta =0$ this
expression reduces to the one found in \cite{AB}. The total current $I$ in
the constriction as a function of the (time-dependent) phase difference $%
\varphi $ calculated from eqs. (\ref{gfr}) and (\ref{Isym}) is shown in
Fig. 3. We see that even relatively small $\zeta $ has strong effect on the
dynamic current-phase relation, suppressing the dc component of the current
and making it more similar to the stationary current-phase relation.

In conclusion, we have developed the microscopic approach to the calculation
of current in short ballistic and diffusive constrictions between the
superconductors with arbitrary quasiparticle spectrum. We used the developed
approach to study the time-dependent current and the dc current in
constrictions between superconductors with the pair-breaking effects, and
also between normal conductors with the proximity-induced superconductivity.

This work was supported in part by the ONR grant \# N00014-95-1-0762,
Russian Fund for Basic Research, grant No. 96-02-18613, and by the U.S.
Civilian Research and Development Foundation under Award No. RP1-165.

\figure{\ DC current-voltage characteristics of a short ballistic
constriction with transparency $D$ between the two superconductors with
pair-breaking effects. The curves are shifted for clarity along the current
axis and illustrate the smearing of the subharmonic gap structure with
increasing strength of the pair-breaking. The values of the pair-breaking
parameter $\zeta$ are (from bottom to top): $\zeta= 0.01,\, 0.1,\, 0.3,\, 1.0
$. The upper curve with $\zeta= 1.0$ corresponds to the regime of gapless
superconductivity. }

\figure{\ DC current-voltage characteristics of a short symmetric $S/NcN/S$
constriction for different values of the constriction transparency $D$. The
curves show the complex subharmonic gap structure associated with the two
energy gaps: $\Delta$ in the $S$ region, and proximity-induced gap $\Delta _g
$ in the $N$ region. Parameter $\gamma _b$ characterizes the transparency of
the $S/N$ interfaces. For discussion see text. }

\figure{\ Dynamic current-phase relation of a short symmetric constriction
between superconductors with pair-breaking effects at low bias voltages,
vanishing temperature, and different values of the pair-breaking parameter. }


\begin{references}
\bibitem{BTK}  T.M. Klapwijk, G.E. Blonder, and M. Tinkham,  Physica B {\bf %
109\&110+C}, 1657 (1982).

\bibitem{AVZ}  S.N. Artemenko, A.F. Volkov, and A.V. Zaitsev, ZhETF {\bf 76}%
, 1816 (1979) [Sov. Phys. - JETP {\bf 49}, 924 (1979)].

\bibitem{Z}  A. V. Zaitsev, ZhETF {\bf 78 }, 221 (1980) [Sov. Phys. -
JETP {\bf 51}, 111 (1980)].

\bibitem{Z1}  A. V. Zaitsev, ZhETF {\bf 86}, 1742 (1984) [Sov. Phys.
- JETP {\bf 59}, 1015 (1984)].

\bibitem{Ar}  G.B. Arnold, J. Low Temp. Phys. {\bf 59}, 143 (1985); {\bf 68}%
, 1 (1987).

\bibitem{GZ}  U. Gunsenheimer and A.D. Zaikin, Phys. Rev. B {\bf 50}, 6317
(1994).

\bibitem{AB1}  D.V. Averin and A. Bardas, Phys. Rev. B {\bf 53}, R1705
(1996).

\bibitem{BSW}  E.N. Bratus, V.S. Shumeiko, and G. Wendin, Phys. Rev. Lett.
{\bf 74}, 2110\ (1995).

\bibitem{AB}  D.V. Averin and A. Bardas, Phys. Rev. Lett. {\bf 75}, 1831
(1995).

\bibitem{Sp}  J.C. Cuevas, A. Mart\'{i}n-Rodero, and A. Levy Yeyati, Phys.
Rev. {\bf 54}, 7366 (1996).

\bibitem{AB2}  A. Bardas and D.V. Averin, cond-mat/9706087.

\bibitem{LO}  A.I. Larkin and Yu.N. Ovchinnikov, ZhETF {\bf 68}, 1915
(1975); {\bf 73}, 299 (1977) [Sov. Phys. - JETP {\bf 41}, 960 (1975); {\bf %
46}, 155 (1977)].

\bibitem{Z2}  A.V. Zaitsev, to be puplished (1997).

\bibitem{Dor}  O.N. Dorokhov, Solid State Commun. {\bf 51}, 381 (1984).

\bibitem{MF}  P.M. Morse and H. Feshbach, {\it Methods of Theoretical
Physics }, (McGraw-Hill, New York, 1953).

\bibitem{Maki}  K. Maki, in: {\it Superconductivity}, Ed. by R. Park,
(Marcel Dekker, New York, 1969), p.1035.

\bibitem{Abr}  A.A. Abrikosov, {\it Fundamentals of the Theory of Metals},
(North Holland, Amsterdam, 1988).

\bibitem{AGK}  B.A. Aminov, A.A. Golubov, and M.Yu. Kupriyanov, Phys. Rev.
B {\bf 53}, 365 (1996).

\bibitem{rem}  Similar model has been considered in \cite{AGK}, but the
calculations of the current there were based on the phenomenological OBTK
approach \cite{OBTK}. Although this approach is valid for fully transparent
constrictions \cite{AB1} it breaks down when the reflection coefficient is
finite. The reason for this is that in \cite{OBTK} MAR is described in terms
of quasiparticle energy distribution function. This does not allow to take
into account quantum mechanical interference between different quasiparticle
trajectories in the energy space which exist at finite reflection
coefficient. Such interference is accounted for in the correct recurrence
relations (\ref{RS}) for quasiparticle wave amplitudes.

\bibitem{OBTK}  M. Octavio, M. Tinkham, G.E. Blonder, and T.M. Klapwijk,
Phys. Rev. {\bf B 27}, 6739 (1983).

\bibitem{VZ}  A.F. Volkov and A.V. Zaitsev, Phys. Rev. B {\bf 53}, 9267
(1996).

\bibitem{FT}  A. Furusaki and M. Tsukada, Physica B {\bf 165\&166}, 1967
(1990).

\bibitem{BH}  C.W.J. Beenakker and H. van Houten, Phys. Rev. Lett. {\bf 66},
3056\ (1991).
\end{references}
\end{document}